\documentclass[prd, aps, preprint, lengthcheck, nofootinbib, floatfix, nolongbibliography]{revtex4-2}
\usepackage[utf8]{inputenc}
\usepackage{graphics,graphicx}
\usepackage{amsmath, amssymb}
\usepackage{multirow}
\usepackage{color}
\usepackage{verbatim}
\usepackage{hyperref}
\usepackage[normalem]{ulem}  
\usepackage{ulem}
\usepackage{color}
\usepackage{cancel}
\usepackage{mathtools}
\usepackage{epstopdf}

\renewcommand\sout{\bgroup\color{blue} \ULdepth=-.5ex \ULset}
\def\slashchar#1{\setbox0=\hbox{$#1$}  
\dimen0=\wd0     
\setbox1=\hbox{/} \dimen1=\wd1  
\ifdim\dimen0>\dimen1   
\rlap{\hbox to \dimen0{\hfil/\hfil}} 
#1     
\else     
\rlap{\hbox to \dimen1{\hfil$#1$\hfil}} 
/      
\fi}

\newcommand{\dd}{\mathrm{d}}
\newcommand{\csavg}{\langle c_s^2 \rangle}
\newcommand{\cs}{c_s^2}
\newcommand{\eps}{\epsilon}

\begin{document}

\title{Average speed of sound in neutron stars}

\date{\today}
\author{Micha\l{} Marczenko}
\email{michal.marczenko@uwr.edu.pl}
\affiliation{Incubator of Scientific Excellence - Centre for Simulations of Superdense Fluids, University of Wroc\l{}aw, plac Maksa Borna 9, PL-50204 Wroc\l{}aw, Poland}

\begin{abstract}
The structure of the dense-matter equation of state is essential for the phenomenology of neutron stars. In this work, I relate the thermodynamic properties to the average speed of sound in the interior of a star. I study the consequences of vanishing of the trace anomaly. In particular, I show that if the trace anomaly vanishes in the centers of maximally massive neutron stars, the speed of sound likely exceeds its conformal value and exhibits nonmonotonic behavior. I also find that the additional assumption of positive definiteness of the trace anomaly naturally induces a local peak of the speed of sound at densities realized in the cores of neutron stars. I also reanalyze the stability condition for hybrid neutron stars. Possible implications for the dense matter equation of state and the phenomenology of neutron stars are also discussed.
\end{abstract}
\maketitle

\section{Introduction}
\label{sec:intro}

Compact stellar objects such as neutron stars (NSs) host matter with densities exceeding the nuclear saturation density ($n_{\rm sat} \simeq0.16~\rm fm^{-3}$). The heaviest NSs have masses exceeding $2~M_\odot$~\cite{Romani:2022jhd}, with densities that reach $6-9~n_{\rm sat}$ at their centers~\cite{Marczenko:2023txe}. This fact makes them one of the densest objects found in the universe. They are therefore unique extraterrestrial laboratories of extremely dense matter. The modern observatories for measuring the masses, radii, and tidal deformabilities of compact objects, the gravitational wave interferometers of the LIGO/Virgo Collaboration (LVC)~\cite{LIGOScientific:2018cki, LIGOScientific:2018hze} and the X-ray observatory Neutron star Interior Composition Explorer (NICER) provide new strong constraints on the mass-radius profile of neutron stars~\cite{Riley:2019yda, Miller:2019cac, Miller:2021qha, Riley:2021pdl}. These stringent constraints suggest that NSs might become veritable probes of extremely dense matter, allowing for a more systematic study of the influence of the formation of different degrees of freedom within the cores of NSs. However, the modeling of NSs requires astrophysical observations and knowledge of the equation of state (EOS) of quantum chromodynamics (QCD), the theory of strong interactions. Thus, the understanding of neutron star physics is at the interface with strong interactions. Major progress in constraining the EOS has recently been made by systematic analyses of recent astrophysical observations of the massive pulsar PSRJ0740+6620~\cite{Cromartie:2019kug, Fonseca:2021wxt, Miller:2021qha, Riley:2021pdl} and PSR J0030+0451~\cite{Miller:2019cac}, and the constraint from the recent GW170817 event~\cite{LIGOScientific:2018cki}, within parametric models of the EoS~(see, e.g.,~\cite{Alford:2013aca, Alford:2017qgh, Annala:2017llu, Annala:2019puf}).

The EOS, $p(\eps)$, provides a relationship between the thermodynamic pressure $p$ and the energy density $\eps$. Knowledge of the EOS can be qualitatively inferred from its derivative with respect to the energy density, which defines the speed of sound. It provides valuable information about the microscopic description of matter. At low temperature, the low-density EOS ($n_B \lesssim 2~n_{\rm sat}$) is reliably provided by chiral effective field theory~\cite{Tews:2018kmu, Hebeler:2013nza, Drischler:2017wtt, Drischler:2020fvz,Drischler:2020hwi,Keller:2022crb,Drischler:2020yad} where $c_s^2$ is found to be below the conformal value of $1/3$. Asymptotically, $c_s^2$ approaches the conformal limit as dictated by the perturbative QCD (pQCD) calculations~\cite{Gorda:2023mkk, Gorda:2021znl, Gorda:2021gha, Kurkela:2016was, Fraga:2013qra, Kurkela:2009gj, Gorda:2018gpy}. However, the density range found in the NSs is not accessible by any of these methods. Knowledge of the EOS between these two extreme limits comes from inferential analyses of astrophysical observables~\cite{Takatsy:2023xzf, Annala:2017llu, Annala:2019puf, Brandes:2022nxa, Brandes:2023hma, Marczenko:2022jhl, Marczenko:2023txe} and effective-model analyses~\cite{Liu:2023ocg, Buballa:2003qv, McLerran:2018hbz, Kovacs:2021ger}. Several suggestions show that the speed of sound exhibits a local maximum at densities realized in the interiors of NSs (see, e.g.,~\cite{Fujimoto:2019hxv, Ecker:2022dlg, Tews:2018kmu}). For example, the tension between the large maximum mass and the small tidal deformability suggests that the pressure increases rapidly at a few times the saturation density, which may imply that the speed of sound is large~\cite{Reed:2019ezm}. However, the structure of the speed of sound remains only hypothetical~\cite{Altiparmak:2022bke, Ecker:2022xxj, Brandes:2022nxa, Brandes:2023hma}. Rapid increase in the speed of sound beyond its conformal value in dense matter is associated with rapid disappearance of the trace anomaly~\cite{Fujimoto:2022ohj}. It should be noted that vanishing of the trace anomaly is a necessary but not sufficient condition for the full restoration of conformal symmetry (see, e.g.,~\cite{Annala:2023cwx}).

EOS-insensitive relations have been found to exist between macroscopic NS observables. They were shown to be useful in estimating several properties of NS (see, e.g.~\cite{Yagi:2016bkt}). Approximately insensitive relations between various astrophysical observables and the ratio of central pressure to central energy density have recently been discovered~\cite{Saes:2021fzr}. This ratio can be interpreted as the average speed of sound inside a star (or the average stiffness of the EOS up to the central energy density of a neutron star)~\cite{Saes:2024xmv}.

This work aims to explore the consequences of the vanishing of the trace anomaly on the EOS. To this end, I express the speed of sound and trace anomaly as functions of the average speed of sound. I find that utilizing low-density constraints and astrophysical constraints provides a lower bound on the maximal value of the speed of sound found in NSs. I also reanalyze the known condition for the stability of NSs after the first-order phase transition in terms of the average speed of sound~\cite{Alford:2013aca}.

This paper is organized as follows. In Sec.~\ref{sec:cs2}, I provide a reinterpretation of basic thermodynamic quantities in terms of the speed of sound and its average. In Sec.~\ref{sec:trace}, I discuss the consequences of the restoration of conformal symmetry for the structure of the speed of sound in neutron stars. In Sec.~\ref{sec:hybrid}, I discuss the condition for stability of NSs after the first-order phase transition. Sec.~\ref{sec:conc} summarizes our results.

\section{Average speed of sound in Neutron Stars}
\label{sec:cs2}
The speed of sound is defined as
\begin{equation}
    c_s^2 \equiv \frac{\dd p}{\dd \eps}\rm.
\end{equation}
Causality and thermodynamic stability imply $0 \leq c_s^2 \leq 1$. A connection between the speed of sound and the trace anomaly $\Delta$ has recently been established~\cite{Fujimoto:2022ohj}:
\begin{equation}\label{eq:cs2}
c_s^2 = \frac{1}{3} - \Delta - \Delta' \textrm,
\end{equation}
where the trace anomaly scaled by the energy density reads
\begin{equation}\label{eq:trace}
    \Delta = \frac{\eps - 3p}{3\eps} = \frac{1}{3} - \frac{p}{\eps} \textrm\\
\end{equation}
and its derivative is
\begin{equation}
    \Delta' = \frac{\dd \Delta}{\dd \ln \eps} = \eps\frac{\dd \Delta}{\dd \eps} \textrm.
\end{equation}
The trace anomaly is bounded between \mbox{$-2/3 \leq \Delta \leq 1/3$} due to thermodynamic stability and causality. The ratio of the central pressure to the central energy density in NSs (Eq.~\eqref{eq:trace}) is an indicator of the average EOS of the matter in the NSs~\cite{Saes:2021fzr}. For example, this is clear in the constant-sound-speed (CSS) model, where $p = \csavg \eps$, with $\csavg = \it const$. Although primitive, the CSS parameterization of EOS has been used successfully in analyses of the NS properties~\cite{Alford:2013aca, Han:2018mtj, Chatziioannou:2019yko, Reed:2019ezm}. However, it should be noted that the speed of sound is not necessarily independent of density and the EOS has a potentially more complex structure~\cite{Tan:2021nat}. Assuming that the NS EOS is smooth and that the pressure is zero at vanishing energy density, the ratio of the pressure to the energy density can be reinterpreted as the average speed of sound as follows~\cite{Saes:2021fzr, Saes:2024xmv}:
\begin{equation}\label{eq:cs2_avg_def}
    \csavg = \frac{1}{\eps} \int\limits_0^{\eps}\dd \eps'\;c_s^2 = \frac{1}{\eps} \int\limits_0^{\eps}\dd \eps'\; \frac{\dd p}{\dd \eps'} = \frac{p}{\eps} \textrm.
\end{equation}
Here, it is understood that $\csavg$ denotes the average speed of sound over the energy interval $\langle 0,\eps \rangle$, and $0 \leq \csavg \leq 1$. I note that approximate EOS-independent relations between the average speed of sound and important astrophysical observables were established in~\cite{Saes:2024xmv}.

Using Eq.~\eqref{eq:cs2_avg_def}, I can rewrite Eqs.~\eqref{eq:cs2} and~\eqref{eq:trace} as follows
\begin{align}
    \Delta &= \frac{1}{3} - \langle c_s^2 \rangle \rm, \label{eq:trace_cs2_avg} \\
    \Delta' &= \langle c_s^2 \rangle - c_s^2 \rm,\label{eq:trace_cs2_avg2}
\end{align}
which express the trace anomaly and its derivative as functions of the speed of sound and its average value. From the above equations, the trace anomaly $\Delta$ is the deviation of the average speed of sound from its conformal value and the derivative $\Delta'$ is the deviation of the speed of sound from its average value. I also note that Eq.~\eqref{eq:trace_cs2_avg2} dictates the condition for the behavior of the trace anomaly; that is, $\Delta$ increases when $\cs$ is smaller than $\csavg$ and decreases when $\cs$ is larger than $\csavg$.

The physical meaning of the trace anomaly is that it is proportional to the rate of change of the number of active degrees of freedom, that is, $\Delta \propto d (P/\mu^4 )/ d \mu$ at zero temperature~\cite{Fujimoto:2022ohj}. Therefore, the negative trace anomaly could indicate the presence of a finite condensate, e.g., quark Cooper pairing in color superconductivity. From $\Delta = 1/3 - \langle c_s^2 \rangle < 0$, such a phase will therefore be characterized by $\langle c_s^2 \rangle > 1/3$.

From Eqs.~\eqref{eq:trace_cs2_avg} and~\eqref{eq:trace_cs2_avg2}, one can see that the restoration of conformal symmetry requires a simultaneous convergence of $\cs$ and $\csavg$ to $1/3$. I also note that in general $\cs$ exceeds the conformal limit before the trace anomaly vanishes. This is clear from Eq.~\eqref{eq:trace_cs2_avg}. The trace anomaly vanishes when $\csavg$ reaches $1/3$, which means that it must exceed the conformal value at smaller energy densities.

A quantity that is closely related to the speed of sound is the polytropic index
\begin{equation}\label{eq:gamma}
    \gamma \equiv \frac{\dd \ln p}{\dd \ln \eps} = \frac{\eps}{p}\cs = \frac{c_s^2}{\csavg} \rm,
\end{equation}
where the last equality holds due to Eq.~\eqref{eq:cs2_avg_def}. The identification of conformal matter was recently associated with a threshold value of the polytropic index $\gamma \simeq 1.75$~\cite{Annala:2019puf}. Interestingly, from Eq.~\eqref{eq:gamma} it is clear that the limiting value of $\gamma$ at high densities is $\gamma \rightarrow 1$, because $\csavg \rightarrow \cs$ as $\eps \rightarrow \infty$. This holds regardless of the high-density limit of the speed of sound. I may exemplify this in the CSS model, where $p = \csavg \eps$, and $\csavg = \it const$. From Eq.~\eqref{eq:gamma}, one gets $\gamma = 1$, which holds regardless of the value of $\csavg$. Another model that vividly shows this result is a simple model with vector interaction~\cite{Fujimoto:2022ohj}, where the energy density and the pressure are $\eps = m_B n_B + C n_B^2$, $p=C n_B^2$, where $m_B$ is the baryon mass, $n_B$ is density, and $C = \it const$, which is connected to the typical interaction strength and the size of the system, and its value is not important for our considerations. Using thermodynamic relations one gets that the polytropic index in this model is given as
\begin{equation}
    \gamma = \frac{Cn_B + m_B}{C n_B + m_B/2 } \rm,
\end{equation}
which approaches unity as the density increases. On the other hand, in the low-density limit, $\gamma \rightarrow 2$. This is consistent with the predictions of the class of relativistic mean-field EOSs for hadronic matter at low density, for which the interactions at low temperature are dominated by the repulsive vector interactions and $\gamma \gtrsim 2$~\cite{Walecka:1974qa,Serot:1984ey,Serot:1997xg}. 

\begin{figure}
    \centering
    \includegraphics[width=\linewidth]{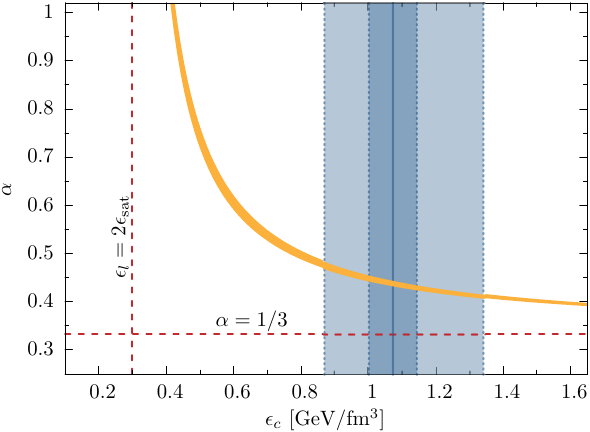}
    \caption{The average speed of sound $\alpha$ as a function of the energy density $\eps_c$. The horizontal, dashed line marks $\alpha = 1/3$. The vertical, dashed line marks the constraint on the energy density $\eps_l = 2\eps_{\rm sat}$ used in this work. The solid, blue line marks the estimate for $\eps_{\rm TOV}$ at the center of maximally massive NSs with the inner (outer) blue regions indicating the $68\%$ ($95\%$) confidence interval from~\cite{Marczenko:2023txe}.}
    \label{fig:cs2_ec}
\end{figure}

\section{Consequences of vanishing trace anomaly}
\label{sec:trace}

The maximal central energy densities [that is, the Tolman-Oppenheimer-Volkoff (TOV) limit] in the NS cores can reach $\eps_{\rm TOV}=1.073^{+0.071(0.267)}_{-0.070(0.202)}~\rm GeV/fm^3$ at 68\% (95\%) confidence level~\cite{Marczenko:2023txe}. The value of the trace anomaly at $\eps_{\rm TOV}$ has recently been estimated to be $\Delta_{\rm TOV} = -0.02\pm0.03(0.1)$ at 68\% (95\%) confidence level~\cite{Marczenko:2023txe}. Interestingly, $\Delta_{\rm TOV}$ is consistent with zero within the uncertainties. Consequently, the centers of the maximally massive NSs may contain matter that is approximately conformal~\cite{Marczenko:2022jhl}. However, the fate of $\Delta$ at densities higher than $\eps_{\rm TOV}$ is unresolved due to the lack of experimental data. One of the key questions is whether or not the trace anomaly remains positive at all densities, especially at densities realized in neutron stars. The answer to this question may have immense consequences in constraining the EOS and determining the NS structure~\cite{Fujimoto:2022ohj}. For example, $\Delta \geq 0$ would constrain the EOS to $p(\eps)\leq \eps/3$. Here, I discuss the implications of vanishing trace anomaly inside NS, in particular the qualitative structure of the speed of sound.

Here, I assume that the trace anomaly vanishes at some finite energy density $\eps_c$, i.e., $\Delta_c = 0$ or equivalently $\csavg_c = 1/3$ [cf.~Eq.~\eqref{eq:trace_cs2_avg}]. Clearly, without prior knowledge of the EOS at low density, one can only infer the average speed of sound at the interval $\langle0,\eps_c\rangle$. On the other hand, if the EOS is known up to $\eps_l < \eps_c$, the average can be separated into two segments, namely
\begin{equation}\label{eq:cs2_integral}
    \csavg_{c} = \csavg_l \frac{\eps_l}{\eps_c} + \alpha \left(1-\frac{\eps_l}{\eps_c} \right) = \frac{1}{3} \textrm,
\end{equation}
where $\csavg_l = p_l/\eps_l$ is the average speed of sound over the energy interval $\langle 0, \eps_l \rangle$ and is assumed to be known, and 
\begin{equation}
\alpha= \frac{1}{\eps_c - \eps_l} \int\limits_{\eps_l}^{\eps_c} \dd \eps \; \cs \textrm,\\
\end{equation}
is the average speed of sound over $\langle \eps_l, \eps_c\rangle$. Eq.~\eqref{eq:cs2_integral} can be solved for $\alpha$:
\begin{equation}\label{eq:alpha}
    \alpha = \frac{\csavg_c - \csavg_l \eps_l/\eps_c}{1-\eps_l/\eps_c} \textrm.
\end{equation}
The knowledge of the low-density EOS allows us to determine the value of $\alpha$ and, thus, estimate the lower bound for the maximum of speed of sound in the cores of NSs. Here, I use the constraint on low-density $\beta$-equilibrium EOS at $\eps_l = 2.0\eps_{\rm sat}$ derived based on chiral effective field theory ($\chi$EFT) interactions to next-to-next-to-next-to-leading (N${}^3$LO) order~\cite{Drischler:2020fvz}. The values are listed in Table~\ref{tab:table1}.

The relation $\alpha(\eps_c)$ is shown in Fig.~\ref{fig:cs2_ec} and is a rapidly decreasing function of energy density. This is because a larger $\alpha$ means that the EOS is stiffer on average. Consequently, the trace anomaly decreases faster and vanishes at smaller energy densities. Causality requires $\csavg \leq 1$, which also yields the maximal value $\alpha = 1$. This condition estimates the lowest value for $\eps_c \simeq 0.42~\rm GeV/fm^3$. $\alpha=1$ corresponds to a CSS parameterization with $p=\eps$ at the interval $\langle \eps_l,\eps_c\rangle$. Although such stiff EOS is rather unrealistic, I have verified that it is still in accordance with the constraint on the tidal deformability $\Lambda_{1.4} = 190^{+390}_{-120}$ of a $1.4~M_\odot$ NS from the GW170817 event measured by the LIGO/Virgo Collaboration (LVC)~\cite{LIGOScientific:2018cki}.

\begin{table}[t!]
    \centering
    \begin{tabular}{|c|c|c|}\hline
        $\eps_l~[\rm GeV/fm^3]$ & $p_l~[\rm GeV/fm^3]$ & $\csavg_{l}$ \\\hline\hline
         0.3 & 0.014 -- 0.023 & 0.047 -- 0.077 \\\hline
    \end{tabular}
    \caption{The constraint on the $\beta$-equilibrium equation of state at $T=0$ based on the N${}^3$LO $\chi$EFT calculations used in this work~\cite{Drischler:2020fvz}. I note that the energy density at saturation is $\eps_{\rm sat} = 0.15~\rm GeV/fm^3$.}
    \label{tab:table1}
\end{table}

To quantitatively estimate the value of $\alpha$ I require that $\eps_c$ has to reproduce the estimated central energy density of maximally massive neutron stars, $\eps_{\rm TOV}=1.073^{+0.267}_{-0.202}~\rm GeV/fm^3$ at $95\%$ confidence level~\cite{Marczenko:2023txe}. This constraint yields $\alpha = 0.41 - 0.49$, which indicates that the speed of sound exceeds the conformal value at energy densities between $\eps_l$ and $\eps_c$. However, I note that sub-conformal equations of state, i.e., $c_s^2 < 1/3$ at $\epsilon < \epsilon_c$, are still possible but unlikely~\cite{Altiparmak:2022bke}.

Eq.~\eqref{eq:cs2_integral} can also be used to estimate the minimal value of $\csavg_c$ for which $\alpha \geq 1/3$. Taking the same $\chi$EFT constraint, I obtain $\csavg_c \geq 0.26$ at the $95\%$ confidence level of $\eps_{\rm TOV}$. This value gives the corresponding threshold $\Delta_c \leq 0.07$. I note that the derived values are, as expected, in agreement with the results obtained in the statistical analyses.

Although the trace anomaly may generally be negative, the conjecture that it may be positive definite at finite density has recently been proposed~\cite{Fujimoto:2022ohj}. If true, it would have consequences for the NS phenomenology. At vanishing chemical potentials and finite temperature, the trace anomaly was determined to be positive from first-principle lattice QCD simulations~\cite{HotQCD:2014kol}. In contrast, the lattice QCD at finite isospin chemical potential produces a negative trace anomaly~\cite{Abbott:2023coj, Brandt:2022hwy}. I also remark that phenomenological nuclear (or hadronic) models of EOS usually render a negative trace anomaly at sufficiently high energy density due to the rapid stiffening of the EOS caused by the contact vector interactions assumed in these models~\cite{Serot:1997xg, Zeldovich:1961sbr, Podkowka:2018gib}. The conjecture of the positivity of the trace anomaly has already gained attention and has been tested, e.g., in model-agnostic approaches~\cite{Ecker:2022dlg, Chatterjee:2023ecc} and theories with modified gravity~\cite{Lope-Oter:2024egz}. Recently, the trace anomaly was extracted from astrophysical mass-radius and red-shift measurements~\cite{Cai:2024oom}.

The assumption that the trace anomaly is non-negative at all densities sets a stringent constraint on the average speed of sound \mbox{$0 \leq \langle c_s^2 \rangle \leq 1/3$} [cf. Eq.~\eqref{eq:trace_cs2_avg}], which otherwise would only be bounded to \mbox{$0 \leq \langle c_s^2 \rangle \leq 1$}.\footnote{The positivity condition of the trace anomaly is different from the sub-conformal speed of sound constraint, which requires \mbox{$\cs \leq 1/3$}.}. As indicated earlier, if the trace anomaly vanishes at $\eps_c$, the speed of sound averages to $\csavg = 1/3$. Under the assumption that $\Delta \geq 0$ at all densities, the average speed of sound cannot exceed $1/3$ and, thus, the speed of sound must develop a local maximum above $c_s^2 = 1/3$ at energy density $\eps \leq \eps_c$. However, It should be noted that it is still possible for $\cs \rightarrow 1/3$ from below and $\Delta \rightarrow 0$ from above asymptotically, so that $\eps_c \rightarrow \infty$ (see, e.g.,~\cite{Fujimoto:2022ohj}).

My results confirm that the large speed of sound is a consequence of the restoration of conformal symmetry in the cores of the heaviest NSs~\cite{Fujimoto:2022ohj}. Moreover, the existence of the peak in the speed of sound could be potentially linked to the positive definiteness of the trace anomaly. I note that the behavior of the EOS discussed in this work, that is, the speed of sound showing a local maximum, is naturally obtained in quarkyonic models~\cite{McLerran:2007qj, Duarte:2021tsx, Kojo:2021ugu, Kojo:2021hqh, Fukushima:2015bda, McLerran:2018hbz, Jeong:2019lhv, Cao:2020byn, Kovensky:2020xif}. Such behavior is in contrast to the transition to quark matter involving the first-order phase transition (FOPT), where the speed of sound is expected to vanish. The quarkyonic matter picture was recently shown to be consistent with the experimental electron scattering data~\cite{Koch:2024zag}. I stress that the results of this section do not rule out any nonmonotonic behaviors in the EOS, such as the FOPT.

\section{Conditions for stable hybrid stars}
\label{sec:hybrid}

\begin{figure}
    \centering
    \includegraphics[width=\linewidth]{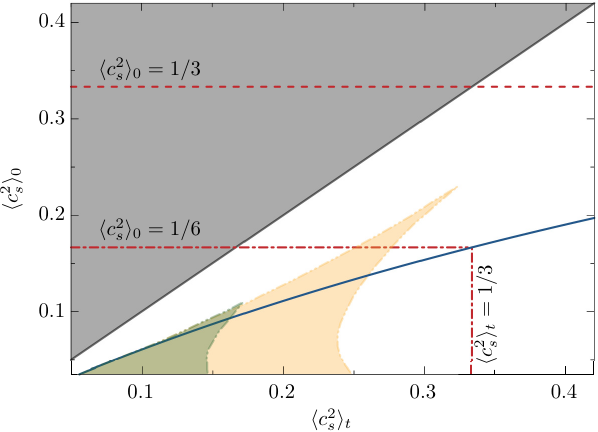}
    \caption{Stability condition from Eq.~\eqref{eq:ns_stable_cs} in terms of the average speed of sound at both sides of the first-order phase transition. The gray area is excluded by the requirement $\csavg_0 < \csavg_t$ (see Sec.~\ref{sec:hybrid} for details). The parameter space that allows for stable (unstable) hybrid NS branches is above (below) the blue, solid line. The green and yellow areas mark the parameter space that allows for stable, but disconnected, hybrid neutron stars based on an exemplary hadronic EOS with $\alpha=1/3$ and $1$, respectively (see Sec.~\ref{sec:hybrid} for details).}
    \label{fig:hybrid}
\end{figure}

In the presence of the FOPT, there is an unstable branch of stars in the mass-radius diagram if the jump in the energy density $\Delta\eps$ at the transition fulfills the following criterion~\cite{seidov1971, kampfer1981}:
\begin{equation}\label{eq:ns_stable}
    \frac{\eps_0}{\eps_t} > \frac{3}{2}\left(1 + \frac{p_t}{\eps_t} \right)\rm,
\end{equation}
where $p_t$ and $\eps_t$ are the pressure and energy density at the phase transition, and $\eps_0 = \eps_t + \Delta \eps$ is the energy density after the transition. Rewriting the above equation in terms of the average speed of sound yields the following condition:
\begin{equation}\label{eq:ns_stable_cs}
    \csavg_0 < \frac{2}{3} \frac{\csavg_t}{1+\csavg_t} \rm,
\end{equation}
where $0 \leq \csavg_t \leq 1$ and $0 \leq \csavg_0 \leq 1$ are the average speeds of sound at the transition and after it, respectively. The range for $\csavg_0$ can be further constrained by noticing that $\csavg_0 < \csavg_t$ at a FOPT. I note that a similar relation for the trace anomaly may also be obtained through Eq.~\eqref{eq:trace_cs2_avg}. Interestingly, to render an unstable NS branch, $\csavg_0 < 1/3$ and, consequently, the trace anomaly after the transition must be positive. Moreover, the requirement of the positivity of the trace anomaly implies that $\csavg_t \leq 1/3$ at the transition and $\csavg_0 \leq 1/6$ after the transition. 

I depict the criterion in Eq.~\eqref{eq:ns_stable_cs} in Fig.~\ref{fig:hybrid}. The combinations of parameters for which the matter becomes unstable after the FOPT are below the blue curve. This region corresponds to roughly $41\%$ of all thermodynamically allowed combinations of $\csavg_t$ and $\csavg_0$. The assumption that the trace anomaly is positive at all densities drastically constrains the allowed parameter space, i.e, $\csavg_t<1/3$ and $\csavg_0 < 1/6$, to roughly $8\%$, only $74\%$ which yield unstable matter after the FOPT.

The mass-radius relations can be classified according to the stability of the branch after the phase transition~\cite{Alford:2013aca}. If an unstable branch is obtained after the FOPT, it may become stable again, leading to a disconnected branch known as the third family of stars (after white dwarfs and neutron stars). However, if the branch remains stable after the FOPT, it may still become unstable and stable again, leading to the same phenomenon. Third families have been widely discussed and studied in specific quark-matter models (e.g.~\cite{Alford:2017qgh, Alvarez-Castillo:2018pve, Christian:2017jni}). To illustrate this, I use the nuclear EOS~\cite{Xia:2022dvw, Xia:2022pja} which is based on a density-dependent relativistic functional theory with covariant density functional DD-ME2, which is an example of a stiff EOS up to a few times the saturation density~\cite{Lalazissis:2005de}. The EOS is then matched with a CSS EOS at $\eps_t$ through a FOPT, such that $\eps(p > p_t) = \eps_0 + \alpha^{-1}(p - p_t)$, where $\alpha$ is the constant value of the speed of sound after the FOPT, i.e., at $p > p_t$. Here, I choose $\alpha = 1$ to maximize the possibility of twin stars. The parameter space that allows for a stable hybrid NS is shown in Fig.~\ref{fig:hybrid} as the yellow region. 

The hybrid-branch region is rather insensitive to the details of the hadronic equation of state~\cite{Alford:2013aca}. I have verified this for the hadronic EOS NL3, which is an example of a soft EOS up to a few times the saturation density~\cite{Lalazissis:1996rd, Xia:2022dvw} (not shown in Fig.~\ref{fig:hybrid}). On the other hand, the hybrid-branch region is highly sensitive to the choice of the equation of state after FOPT~\cite{Alford:2013aca}. For EOSs softer after FOPT (i.e. smaller $\alpha$), the region shrinks because the matter is not able to sustain gravitational collapse. However, the qualitative structure does not depend on the choice of $\alpha$. I illustrate this in Fig.~\ref{fig:hybrid} for the DD-ME2 EOS and $\alpha=1/3$.

The hybrid NS branches are obtained for EOSs with rather small values of the average speeds of sound at both sides of the FOPT. Notably, the parameter space that renders the hybrid NS branches is bounded by $\csavg_t < 1/3$ and $\csavg_0 < 1/3$. This constraint corresponds to a positive trace anomaly on both sides of FOPT. Interestingly, the disconnected hybrid branches lie in the region constrained by the positivity of the trace anomaly.

\section{Conclusion}
\label{sec:conc}

In this work, I investigated the cold and dense equation of state under neutron star conditions. By general thermodynamic considerations, I have provided a reinterpretation of the key thermodynamic quantities in terms of the speed of sound and its average value. I utilized these relations in connection with the results of effective chiral field theory and recent estimates for the values of the central energy density of the maximally massive NSs, $\eps_c \simeq 1~\rm GeV/fm^3$, at which the trace anomaly $\Delta \simeq 0$. I demonstrated that the restoration of conformal symmetry in the centers of the heaviest NSs may imply a large speed of sound above its conformal value at energy densities smaller than $\eps_c$. I also studied the consequences of an additional assumption of the positivity of the trace anomaly. I have shown that it implies a local peak in the speed of sound at densities reachable in the neutron star cores. Lastly, I have reformulated the stability criterion for the first-order phase transition in terms of the average speed of sound and discussed the possibility of a stable hybrid branch (i.e. third family of NSs) in the mass-radius diagram. This allowed us to draw additional conclusions. I found that to render an unstable NS branch after the first-order phase transition, the average speed of sound in the NSs must be less than 1/3. Moreover, the assumption of the positivity of the trace anomaly drastically constrains the allowed parameter space. Further development in constraining the dense matter EOS from first-principle calculations and astrophysical constraints will be essential in determining the structure of the EOS at densities relevant to NS phenomenology.

\section*{Acknowledgments}
The author thanks for constructive remarks from Pasi Huovinen and David Edwin Alvarez-Castillo, and acknowledges fruitful discussions with Larry McLerran, Krzysztof Redlich, and Chihiro Sasaki. This work is supported partly by the program Excellence Initiative–Research University of the University of Wroc\l{}aw of the Ministry of Education and Science.

\bibliography{biblio}

\end{document}